\documentclass[prx,aps,twocolumn,amsmath,footinbib]{revtex4-2}
\usepackage{amssymb}
\usepackage{graphicx}
\usepackage[dvipsnames]{xcolor}
\definecolor{myblue}{RGB}{29, 113, 184}
\definecolor{myred}{RGB}{217, 61, 61}
\usepackage{hyperref}
\hypersetup{
	colorlinks=true,
	linkcolor=myred,
	filecolor=magenta,      
	urlcolor=myblue,citecolor=myblue}
\usepackage{cleveref}
\crefname{figure}{fig.}{figs.}   
\Crefname{figure}{Fig.}{Figs.}   

\crefname{equation}{eq.}{eqs.}   
\Crefname{equation}{Eq.}{Eqs.}   

\usepackage{color} 

\usepackage[caption=false]{subfig}
\begin{document}

\title{Learning Minimal Representations of Many-Body Physics\\ from Snapshots of a Quantum Simulator}

\author{Frederik M{\o}ller\textsuperscript{1,2}}
\author{Gabriel Fern\'andez-Fern\'andez\textsuperscript{3}}
\author{Thomas Schweigler\textsuperscript{2}}
\author{Paulin de Schoulepnikoff\textsuperscript{4}}
\author{J\"{o}rg Schmiedmayer\textsuperscript{2}}
\author{Gorka Mu\~noz-Gil\textsuperscript{4}} 

\affiliation{
\textsuperscript{ 1} Institute of Science and Technology Austria, Am Campus 1, 3400 Klosterneuburg, Austria \\
\textsuperscript{ 2} Vienna Center for Quantum Science and Technology (VCQ), Atominstitut, TU Wien, Vienna, Austria\\
\textsuperscript{ 3} ICFO – Institut de Ci\`encies Fot\`oniques, The Barcelona Institute of Science and Technology,
Av. Carl Friedrich Gauss 3, 08860 Castelldefels (Barcelona), Spain\\
\textsuperscript{ 4} University of Innsbruck, Department for Theoretical Physics, Technikerstr. 21a, A-6020 Innsbruck, Austria
}


\begin{abstract} 
Analog quantum simulators provide access to many-body dynamics beyond the reach of classical computation. However, extracting physical insights from experimental data is often hindered by measurement noise, limited observables, and incomplete knowledge of the underlying microscopic model. Here, we develop a machine learning approach based on a variational autoencoder (VAE) to analyze interference measurements of tunnel-coupled one-dimensional Bose gases, which realize the sine–Gordon quantum field theory. Trained in an unsupervised manner, the VAE learns a minimal latent representation that strongly correlates with the equilibrium control parameter of the system.
Applied to non-equilibrium protocols, the latent space uncovers signatures of frozen-in solitons following rapid cooling, and reveals anomalous post-quench dynamics not captured by conventional correlation-based methods. These results demonstrate that generative models can extract physically interpretable variables directly from noisy and sparse experimental data, providing complementary probes of equilibrium and non-equilibrium physics in quantum simulators. More broadly, our work highlights how machine learning can supplement established field-theoretical techniques, paving the way for scalable, data-driven discovery in quantum many-body systems.
\end{abstract} 

\maketitle 

Quantum analog simulators provide a powerful platform to investigate complex many-body systems by engineering physical setups whose intrinsic dynamics implement the desired model Hamiltonian~\cite{PhysRevLett.81.3108, RevModPhys.86.153,Gross2023-pd}.
In these experiments, engineered quantum systems evolve naturally under the governing physical laws, and relevant observables are extracted through measurement.
This approach has been fruitfully applied across diverse platforms, from ultracold and neutral atoms~\cite{Bloch2012, Ebadi2021, Bluvstein2022} to trapped ions~\cite{Blatt2012, RevModPhys.93.025001} and superconducting circuits~\cite{Houck2012}, enabling insight into strongly correlated phases, non-equilibrium dynamics, and topological phenomena beyond the reach of classical computation~\cite{Daley2022}.
In terms of continuous systems, a particularly well-studied example is the realization of one-dimensional (1D) superfluids, where coherent condensates or coupled arrays serve as testbeds for low-energy quantum field theories describing collective fluctuations~\cite{RevModPhys.83.1405, doi:10.1126/science.aal3837, schweigler2017experimental, Tajik2023, doi:10.1073/pnas.2301287120, doi:10.1126/science.aal3837, PhysRevX.15.010501}.

The read-out from quantum simulators, however, poses unique challenges compared to digital simulations.
For instance, measurement constraints limit access to only a subset of observables~\cite{PhysRevLett.121.080406}, often further degraded by finite imaging resolution and noise.
Consequently, extracting comprehensive information demands both tailored measurement protocol and advanced data analysis methods~\cite{PhysRevLett.123.063603, Gluza2020, PhysRevResearch.7.L022031, PhysRevLett.133.250403}.
These include full counting statistics~\cite{doi:10.1126/science.1224953, Wienand2024}, higher-order correlation functions~\cite{doi:10.1126/science.1257026, schweigler2017experimental, PhysRevX.10.011020}, and increasingly, machine learning (ML) techniques that leverage modern statistical inference to infer hidden degrees of freedom and dynamical rules from experimental trajectories~\cite{Wang2017, Bohrdt2019, PhysRevResearch.6.043284, Lange2023adaptivequantum, schüttelkopf2024characterisingtransportquantumgas, Lange2025transformerneural, f58h-zxs3}.

Another central difficulty arises when the simulator’s effective Hamiltonian or noise processes are not fully known a priori or when experimental datasets are limited, making direct parameter estimation unreliable or biased. Therefore, methods that are agnostic to detailed microscopic assumptions yet sufficiently expressive and interpretable to reveal underlying physics are highly desirable~\cite{Interpretable_Wetzel_2025,Machine_Miles_2023, PhysRevB.109.075152}.

In this work, we address this challenge by employing a custom variational autoencoder (VAE) trained on the relative phase field measurements of a quantum simulator composed of two tunnel-coupled 1D ultracold Bose gases.
The proposed VAE, an unsupervised generative model~\cite{kingma2013auto} equipped with an autoregressive architecture~\cite{van2016wavenet}, identifies a minimal parametric representation of the stochastic process governing phase fluctuations of the system without relying on prior knowledge of the Hamiltonian or noise sources.
We demonstrate that the representation learned by the model strongly correlates with physical parameters controlling the effective sine–Gordon field theory realized in the experiment.
Furthermore, the model’s representation provides a sensitive probe of non-equilibrium phenomena, including the detection of topological soliton defects and the characterization of post-quench dynamics where conventional observables fail to discriminate equilibrium from non-equilibrium states.

\begin{figure*}
    \centering
    \includegraphics[width=\textwidth]{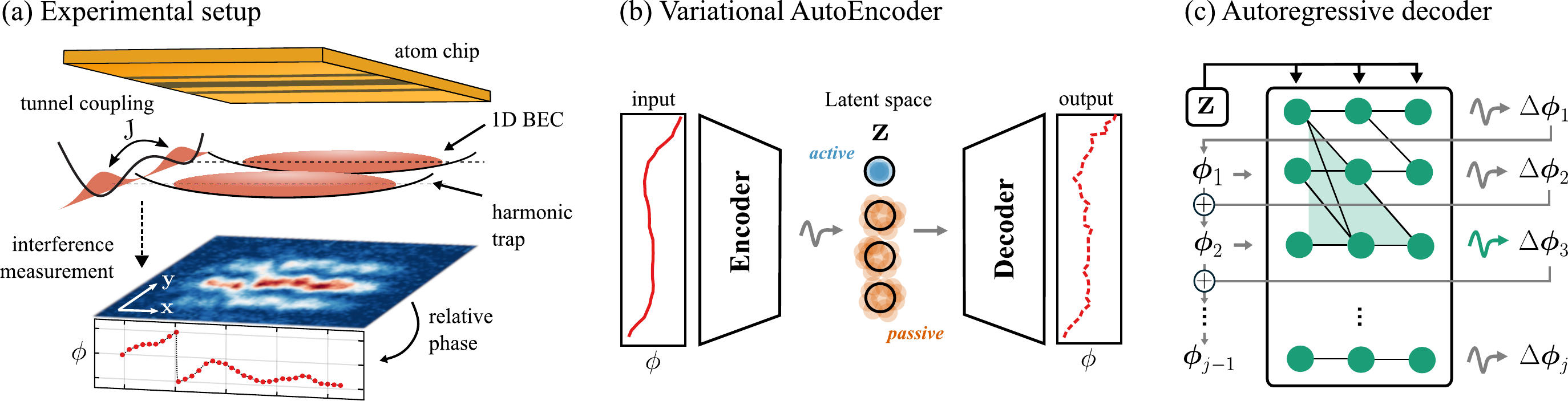}
    
    \caption{\label{fig:fig1} \textbf{Experimental and computational pipeline.}
    (a) Two quasi–one-dimensional superfluids in a magnetic double-well created by an atom chip with tunable tunnel coupling $J$. The relative phase $\phi$ is extracted from absorption imaging of the matter-wave interference pattern.
    (b) The encoder maps each phase trajectory to a lower-dimensional latent vector $\mathbf{z}$. The model compresses the input into a minimal set of neurons (active), effectively suppressing non-contributing ones (passive). The decoder reconstructs then the phase trajectory from $\mathbf{z}$.
    (c) The decoder predicts the conditional probability of the next phase increment $\Delta \phi_{j}$ based on $\mathbf{z}$ and the previous phases $\phi_{i<j}$.
    Its autoregressive nature ensures causal prediction (green shaded area)
    }
\end{figure*}

\subsection*{Experimental realization of the sine–Gordon quantum simulator}

We study two quasi–one-dimensional, tunnel-coupled $^{87}$Rb superfluids prepared in a magnetic double-well potential created by an atom chip (see \Cref{fig:fig1} and also Refs.~\cite{schweigler2017experimental, Schweigler2019} for further details).
At low energies, their relative degrees of freedom are captured by the sine–Gordon Hamiltonian~\cite{PhysRevB.75.174511}:
\begin{equation}
    H_{\mathrm{sG}}=\int_0^L \mathrm{d} x\left[g\  \delta \rho^2+\frac{\hbar^2 n}{4 m}\left(\partial_x \phi\right)^2-2 \hbar J n \cos (\phi)\right] \, ,
\end{equation} 
where $\delta\rho(x)$ and $\phi(x)$ are canonically conjugate relative density and phase fields, $m$ is the atomic mass, $g$ the 1D contact-interaction parameter, $n(x)$ the linear density, which varies slowly due to the weak axial confinement, and $J$ the tunnel coupling.
The relative phase field $\phi(x)$ is accessed through matter-wave interference~\cite{Schumm2005}, recorded with a CCD camera of pixel size $\Delta x \approx 2\mu\mathrm{m}$, and henceforth treated as a discrete quantity $\phi_j$, which we refer to as a \emph{phase trajectory}.
This destructive measurement requires repeated preparation and imaging, causing that each realization yields a single sample trajectory subject to shot-to-shot fluctuations of both temperature and atom number arising from technical imperfections. 

The sine–Gordon model provides a unifying description of one-dimensional systems: it can be viewed as a Luttinger liquid~\cite{tomonaga1950remarks, luttinger1963exactly} perturbed by a cosine interaction potential, admits solitonic excitations familiar from classical nonlinear dynamics~\cite{FADDEEV19781}, and has long served as a minimal quantum field theory relevant in condensed matter~\cite{RevModPhys.51.659}, gauge theories~\cite{COLEMAN1976239}, and cosmology~\cite{Cuevas-Maraver2014}.
In our setting, the strength of the cosine interaction is tunable by controlling the double-well separation, and thus $J$, through radio-wave dressing, thereby enabling simulation of various interaction regimes.

In thermal equilibrium, the coherence properties are governed by the dimensionless coupling $Q = \lambda_T/l_J$, with $l_J=\sqrt{\hbar/(4 m J)}$ the Josephson length and  $\lambda_T = 2 \hbar^2 n/(m k_{\mathrm{B}} T)$ the thermal coherence length~\cite{PhysRevA.87.013629}. 
Systems with equal $Q$ exhibit identical behavior when expressed in units of $\lambda_T$, so $Q$ fully controls the phase statistics. 
While not directly measurable, $Q$ is related to the coherence factor $\langle \cos \phi \rangle$ (see \Cref{app:imaging}). 
In the strong-coupling regime ($Q \gg 1$), tunneling locks the relative phase near $\phi \approx 0$, yielding $\langle \cos \phi \rangle \approx 1$ and long-range correlations. 
When thermal fluctuations dominate ($Q \lesssim 1$), the relative phase is randomized, $\langle \cos \phi \rangle \to 0$, and correlations decay exponentially over $\lambda_T$~\cite{PhysRevLett.106.020407}.

Using the transfer matrix formalism~\cite{PhysRevB.11.3535, PhysRevB.22.477}, for homogeneous condensates in thermal equilibrium, the fluctuating thermal phase field of the sine--Gordon model can be expressed using the stochastic It\^{o} equation~\cite{PhysRevA.98.023613, risken1989fpe}, where position takes the role usually taken by time:
\begin{equation}
    \mathrm{d}\phi = A[\phi] \mathrm{d}x + \sqrt{2 D \mathrm{d}x} \: \mathcal{N}(0, 1) \; ,
    \label{eq:ito}
\end{equation}
where $A$ is the deterministic $Q$-dependent drift term obtained by diagonalizing the transfer operator (see \Cref{app:ito}), and $D =  2 \lambda_{\mathrm{T}}^{-1}$ is the diffusion constant.
Heuristically, the drift term encodes the restoring force of the cosine potential through the tunnel coupling, while $D$ encodes the thermal randomization of the phase.
In practice, the measured phase trajectories deviate from the ideal It\^{o} process of \Cref{eq:ito}, as the finite resolution and coarse-graining of the imaging system smooth the underlying fluctuations.
Further, the harmonic confinement of the magnetic trap introduces a weak density variation; we focus on the central $\approx 70~\mu\mathrm{m}$ (35 pixels) of relative-phase images, where the linear density $n$ varies between 70-140~$\mu\mathrm{m}^{-1}$.

\subsection*{Variational autoencoder architecture}
To extract low-dimensional, physically interpretable representations from our experimental data, we employed a variational autoencoder (VAE) architecture~\cite{kingma2013auto}, which has proven effective in capturing the essential features of stochastic physical dynamics~\cite{fernandez2024learning, de2025interpretable}. A VAE comprises three main components: an encoder, a latent space, and a decoder (see \Cref{fig:fig1}). The encoder, built here via a stack of convolutional and feed-forward layers, compresses each input phase trajectory into a small set of latent variables. Such latent space is parameterized by a multivariate Gaussian distribution, where the mean $\mu_i$ and variance $\sigma_i$ of each latent variable $z_i$ are predicted by the encoder. Samples $\mathbf{z}$ drawn from this distribution are then passed to the decoder.

The decoder reconstructs the input from the latent variables. Importantly, as shown in Refs.~\cite{fernandez2024learning, de2025interpretable}, the decoder must faithfully reproduce the statistical properties of the training data to yield a meaningful latent representation. Conceptually, the decoder learns the physical mapping from the system’s underlying degrees of freedom to the experimentally observed trajectories. To achieve this, we design it to model the conditional probability of the next phase increment given the current phase history and the latent variables,
\begin{equation}
p_\theta(\Delta \phi_{j+1}\mid \phi_j, \phi_{j-1},...,\mathbf{z})\, ,
\end{equation}
where $\theta$ denotes the set of trainable parameters of the decoder. In this way, the model is explicitly encoding the stochastic nature of the dynamics in its output distribution. In our system, we expect the next increment $\Delta\phi_{j+1}$ to depend directly on the present phase $\phi_j$, given the deterministic drift term in the effective It\^{o} process is a function of the current phase (see \Cref{eq:ito}).
Consequently, we adopt an autoregressive structure conditioned on past phase values, rather than past increments. In practice, the conditional distribution is parameterized using a WaveNet architecture~\cite{van2016wavenet}, which employs dilated convolutions throughout the previous phase points (shaded region in \Cref{fig:fig1}c) to correctly capture correlations of the data.

The VAE is trained by maximizing the Evidence Lower Bound (ELBO)~\cite{kingma2013auto}. For a dataset of $N$ phase trajectories $\boldsymbol{\phi}^{(n)}$, the loss function reads as
\begin{equation} \label{eq:loss_VAE}
    \mathcal{L} = \frac{1}{N}\sum_{n=1}^{N}\mathbb{E}_q[\log p_\theta(\boldsymbol{\phi}^{(n)}|\mathbf{z})] - \beta \mathrm{KL}[q_{\tilde{\theta}}(\mathbf{z}|\boldsymbol{\phi}^{(n)})||p(\mathbf{z})].
\end{equation}
This expression contains two contributions: (i) a reconstruction loss, which maximizes the likelihood of the input data under the decoder $p_\theta$; and (ii) a regularization term, which aligns the latent distribution $q_{\tilde{\theta}}(z_i)$ with the standard Gaussian prior $p(z_i)= \mathcal{N}(0,1)$ through the Kullback–Leibler (KL) divergence. Here, $q_{\tilde{\theta}}$ denotes the encoder with trainable parameters $\tilde{\theta}$. The weighting factor $\beta$ balances the relative strength of the two terms.
Following~\cite{Chen_Isolating_2018}, we adopt here the Total Correlation VAE formulation, which further encourages conditional independence among latent variables (see \Cref{app:ml}), typically leading to more interpretable latent spaces.

In practice, the competition between the reconstruction and regularization terms in \Cref{eq:loss_VAE} drives the model to encode only the minimal set of latent neurons required to accurately reconstruct the data. Those neurons that do not meaningfully contribute to the reconstruction collapse to the prior, i.e. $\mu_i\rightarrow0$ and $\sigma_i\rightarrow1$, becoming \emph{passive} neurons. This process effectively filters out noise and redundancies in the representation. When the dataset is governed by a small number of hidden degrees of freedom—as in many physical systems—the \emph{active} latent variables naturally align with these underlying parameters~\cite{iten2020discovering,nautrup2022operationally}, providing a minimal and physically interpretable coordinate system for the observed dynamics. As we will see below, this mechanism enables the VAE to distill complex, noisy phase trajectories into a reduced description that correlates strongly with the control parameters of the sine–Gordon model.

\section*{Results}

\subsection*{Training VAEs on experimental interference measurements}
We train the autoregressive VAE via Ref.~\cite{spivae_repo} on experimental equilibrium phase trajectories obtained across several tunnel couplings $J$~\cite{schweigler2017experimental}. 
The datasets contain around $10^3$ experimental trajectories per $J$. Starting from a six-dimensional latent space, the VAE robustly collapses all but one neurons to the uninformed prior. This is shown in \Cref{fig:latent_analysis}a, where, for 20 different training initializations, the variances $\sigma_i$ of the collapsed neurons converge to one (orange lines) while the remaining active neuron $z_a$ remains at values $\sigma_a\ll 1$ (blue line). Hence, for the rest of this work, we will only analyze the active neuron and also consider that $z_a \leftarrow \mu_a$ due to its small variance.
This result identifies a single dominant degree of freedom in the input data, consistent with the experiment effectively varying one control parameter (the coupling $J$) and with the sine–Gordon equilibrium process being governed by the single dimensionless parameter $Q$. 
That only one neuron is active despite experimental imperfections and imaging noise highlights the ability of the VAE to robustly learn the physical model underlying the input data even in such challenging conditions.
Indeed, when analyzing the values of the active latent neuron for the training experimental trajectories, we observe a broad distribution for all $J$s, reflecting shot-to-shot parameter fluctuations of temperature and linear density.

\subsection*{Interpreting the physical model learned in the latent space}

\begin{figure*}
    \centering
    \includegraphics{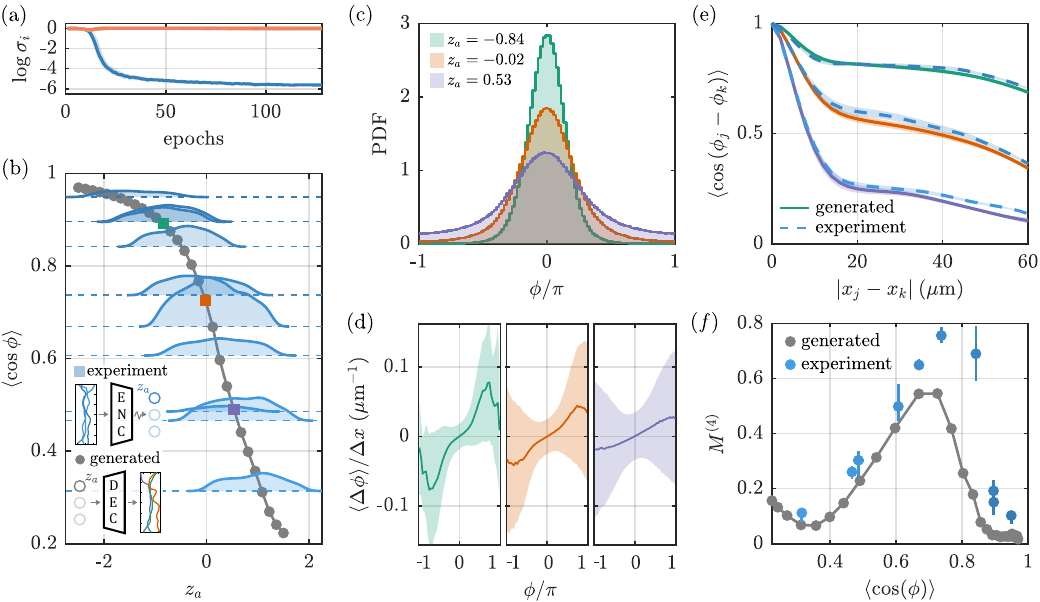}
    
    \caption{\label{fig:latent_analysis}
    \textbf{Interpreting the latent space of the trained VAE.}
    (a) Median variance of active (blue) and passive (orange) latent neurons over 20 trained models.
    (b) Coherence factor $\langle\cos\phi\rangle$ computed from trajectories generated at fixed values of the active neuron $z_a$ (gray points) and from subsets of the experimental training data (blue horizontal markers). For each subset, the distribution of activation obtained by passing the data through the encoder is shown vertically; the distribution area reflects the varying number of experimental samples.
    (c) Histograms of generated phases $\phi$ for three representative $z_a$ values.
    (d) Mean (solid) and standard deviation (shaded) of generated phase increments $\Delta\phi_{j+1}$ as a function of $\phi_j$.
    (e) Circular two-point correlations as function of distance for the same three representative $z_a$ values and for experimental subsets with matching coherence factors. Shaded areas indicate standard error.
    (f) Connected fourth moment $M^{(4)}$ versus coherence factor for generated and experimental data; experimental error bars indicate 80\% bootstrap confidence intervals.
    }
\end{figure*}

To interpret the physics encoded in the active latent neuron, we use the generative capabilities of the model and sample from the decoder ensembles of trajectories conditioned on a wide range of latent values $z_a$, keeping in all cases the value of the collapsed neurons to zero.

As shown in \Cref{fig:latent_analysis}b (gray line), the coherence factor $\langle\cos\phi\rangle$ computed from the generated samples depends continuously on $z_a$, even when the original training data only covered, in principle, a discrete set of values (blue horizontal lines). Notably, $z_a$–dependence resembles the theoretical relation between $\langle\cos\phi\rangle$ and $Q$, up to a sign and an overall scale (see \Cref{fig:coherence_Ito}; a feature that was not enforced but autonomously learned by the VAE.

To further assess the relation between the active neuron and the generated phases' properties, we analyze their distribution at representative values of $z_a$ (marked by colored points in \Cref{fig:latent_analysis}b).
\Cref{fig:latent_analysis}c shows the expected behavior: at negative $z_a$ (related to high coherence) phases concentrate near $\phi=0$, while at larger $z_a$ the distribution broadens and weight near $\pm\pi$ appears, indicating trajectories that cross into neighboring vacua of the interaction potential.

The decoder output constitutes an effective It\^{o}-type process.
To assess the nature of such process and its relation to the It\^{o} process in \Cref{eq:ito}, we analyze the generated increments $\Delta\phi$ as a function of the current phase $\phi$ (mod $2\pi$).
Importantly, this analysis is made possible by the model’s ability to generate a large number of trajectories, providing sufficient statistics that the limited experimental dataset alone could not offer.
As shown in \Cref{fig:latent_analysis}d, the mean increment produces a restoring drift comparable to \Cref{eq:ito}, with its strength increasing for negative $z_a$.
Interestingly, the diffusion term, emerging here from the increment variances (shaded areas), shows a notable $\phi$-dependence, which is absent in the ideal It\^{o} process.
We attribute this dependence to the experimental imaging process, effectively mixing the terms of the underlying process.
Indeed, simulations of the ideal process accounting for finite imaging resolution and coarse-graining~\cite{Schweigler2019, PhysRevA.104.043305} reproduce this effect qualitatively (see \cref{app:ito}).

An essential feature of this process is the emergence of long-range phase order associated with finite tunnel coupling, which is directly reflected in the two-point correlation function~\cite{PhysRevLett.106.020407}.
As illustrated in Fig.~\ref{fig:latent_analysis}e, these correlations decay exponentially at short distances, then approach a nonzero plateau set by $Q$.
At larger separations, a slow residual decay appears, reflecting the weak density inhomogeneity of the trapped condensates. 
Phase trajectories generated by the decoder reproduce these behaviors, confirming that the latent representation captures both short-range thermal fluctuations and long-range phase locking.

Beyond such two-point correlations, a defining feature of this system is the presence of high order non-Gaussian phase correlations between the coupled condensates~\cite{schweigler2017experimental, Schweigler2021}.
To quantify these correlations, we compute the normalized connected fourth moment
\begin{equation}
\label{eq:m4}
    M^{(4)}= \frac{\sum_{\mathbf{j}}\left|\left\langle \left( \phi_{j_1} - \phi_0 \right) \ldots \left( \phi_{j_4} - \phi_0 \right) \right\rangle _{c} \right|}{\sum_{\mathbf{j}}\left|\left\langle \left( \phi_{j_1} - \phi_0 \right) \ldots \left( \phi_{j_4} - \phi_0 \right) \right\rangle \right|}  \; ,
\end{equation}
where $\mathbf{j} = \{ j_1, j_2, j_3, j_4 \}$ and $j_1 \leq j_2 \leq j_3 \leq j_4$, $\phi_0$ indicates a fixed reference point, and $\langle\cdot\rangle_c$ the connected correlator.
In \Cref{fig:latent_analysis}f, we plot $M^{(4)}$ versus $\langle\cos\phi\rangle$  for both the experimental phase trajectories (blue) and trajectories generated by the VAE for the same range of $z_a$  shown in \Cref{fig:latent_analysis}b. Remarkably, both ensembles showcase the same qualitative behavior:
In the low-coherence limit, $\langle \cos\phi\rangle \to 0$, the cosine interaction potential in the sine–Gordon Hamiltonian becomes negligible and the system reduces to a quadratic Tomonaga–Luttinger liquid~\cite{tomonaga1950remarks, luttinger1963exactly}.
Hence, higher-order connected correlations, and thus $M^{(4)}$, vanish.
At the opposite extreme, $\langle \cos\phi\rangle \to 1$, the tunnel coupling strongly locks the relative phase near $\phi=0$. In this regime, the cosine potential can be expanded to quadratic order, yielding the massive Klein–Gordon model~\cite{PhysRevA.68.053609}, which is again Gaussian and produces vanishing $M^{(4)}$.
Only at intermediate coherence factors, where the cosine potential is neither negligible nor purely harmonic, do genuine non-Gaussian states contribute, leading to finite values of $M^{(4)}$.

Interestingly, the generated trajectories produce systematically lower $M^{(4)}$ than experiment.
We attribute such behavior to the sampling of the generated trajectories being performed at fixed $z$, whereas each experimental data set spans a broad range of latent activations (see \Cref{fig:latent_analysis}b), reflecting variation of physical parameters within the set.

\subsection*{Detecting topological defects via latent representations}

\begin{figure}
    \centering
    \includegraphics{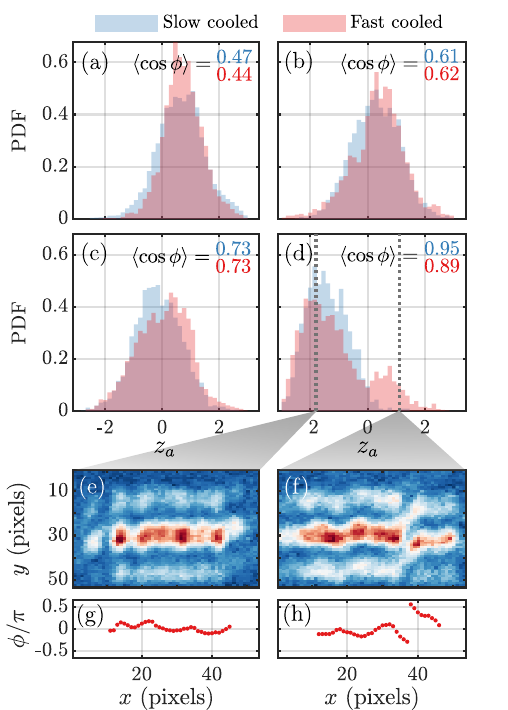}
    
    \caption{\label{fig:fast_cooled}
    \textbf{Latent-space discrimination of solitonic defects in fast-cooled data.}
    (a–d) Activation histograms of the active latent neuron for equilibrium (slow-cooled) and fast-cooled trajectories with matched coherence factors.
    (e–h) Representative phase trajectories and corresponding absorption images from the negative-activation (equilibrium-like) and positive-activation (solitonic excitation) peaks.
    }
\end{figure}

Having established that the VAE faithfully reproduces the equilibrium, measurement-limited sine–Gordon statistics, we next use the previously trained model as a discriminative probe for non-equilibrium states.
For this purpose, we consider a second experimental dataset in which the final cooling ramp is ten times faster than in the equilibrium runs~\cite{schweigler2017experimental}.
Such rapid cooling can freeze in domain regions that settle into different minima of the cosine interaction potential.
Hence, the phase must wind by $2\pi$ across the domain boundaries, creating topological solitons of the sine–Gordon model.
In this regime the solitons are static on experimental timescales and the field fluctuates around them, representing excitations above topologically distinct “false” vacua~\cite{PhysRevD.16.1248}.
Although rare at strong coupling, even a few of such defects markedly alter higher-order moments~\cite{schweigler2017experimental}.

We input the fast-cooled trajectories to the equilibrium-trained VAE's encoder and compare, in \Cref{fig:fast_cooled}a-d, the resulting latent activation histograms (red) to those from equilibrium data with similar coherence factors (blue).
For weak coupling (\Cref{fig:fast_cooled}a-c), the two datasets are nearly indistinguishable, as thermally excited solitons are already common in equilibrium.
However, at stronger coupling (\Cref{fig:fast_cooled}d), the fast-cooled data develop a clear second peak at positive activations in addition to the equilibrium-like negative peak.
As analyzed in \Cref{fig:latent_analysis}c, negative activations correspond to trajectories localized near the main vacuum $\phi \approx 0$, whereas positive activations signal large phase increments, which, at strong coupling, is associated with solitonic defects.
To confirm such behavior, we examine samples assigned to both peaks, both from its original absorption images (\Cref{fig:fast_cooled}e-f) and their corresponding phase trajectories (\Cref{fig:fast_cooled}g-h).
As shown, while samples with negative $z_a$ stay in the main vacuum, positive values lead to trajectories exhibiting localized $2\pi$ rotations that connect neighboring vacua, with their absorption images revealing corresponding phase slips in the interference fringes.
Thus, the encoder trained solely on equilibrium data can identify topologically non-trivial excitations in the quantum simulator.

\subsection*{Probing quench dynamics beyond equilibrium field theory}

\begin{figure}
    \centering
    \includegraphics{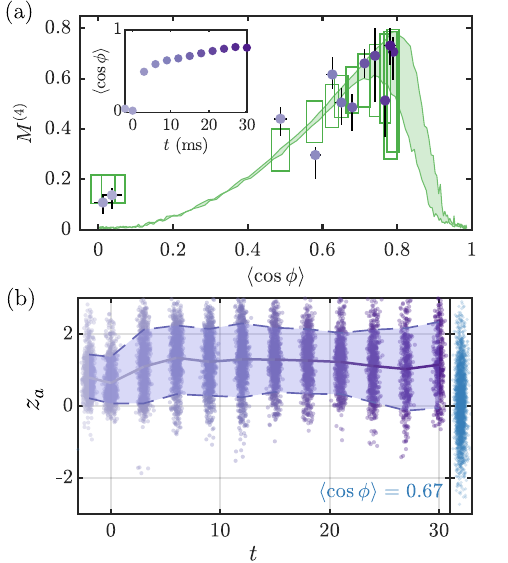}
    
    \caption{\label{fig:quench}
    \textbf{Latent-space analysis of quench dynamics.}
    (a) Fourth connected moment $M^{(4)}$ for post-quench experimental trajectories, compared to equilibrium sine–Gordon predictions from It\^{o}-process sampling with imaging effects. The shaded area spans $\lambda_T = 20\:\mu\mathrm{m}$ (top) to $\lambda_T = 30\:\mu\mathrm{m}$ (bottom). Rectangles show theory for $\lambda_T = 20\:\mu\mathrm{m}$ with the same number of samples as in the experiment; widths and heights indicate 80$\%$ confidence intervals. 
    Inset shows measured coherence factor as function of time.
    (b) Latent neuron activation versus evolution time $t$ for the quench data, with solid line and shaded area giving the median and 80$\%$ confidence interval. Blue distribution: equilibrium subset with coherence factor matched to the final quench value.
}
\end{figure}

Finally, we turn to a third experimental protocol involving a sudden quench of the tunnel coupling~\cite{Schweigler2019, PhysRevLett.120.173601}.
In this sequence, the system is first prepared in a fully uncoupled state $J = 0$.
The magnetic trapping potential is then rapidly quenched to a configuration with finite coupling $J$, and measurements are taken at a series of evolution times $t$ after the quench.

Following the quench, the coherence factor immediately increases sharply, then continues to grow gradually (see \Cref{fig:quench}a inset), qualitatively similar to theoretical predictions~\cite{PhysRevLett.110.090404}.
At first glance, conventional analysis based on the connected fourth moment $M^{(4)}$ (\Cref{eq:m4}) yields a somewhat surprising result: the post-quench trajectories appear to follow the same functional dependence as equilibrium sine–Gordon ensembles (see \Cref{app:imaging}), with an effective coupling strength that evolves in time.
Taken at face value, this would suggest that the system rapidly equilibrates to a thermal state despite the substantial energy injected by the quench.
Yet such an interpretation is counterintuitive.
Starting from $J=0$ means the relative phase is initially random across $[-\pi,\pi)$, whereby quenching to finite interaction excites high-energy modes, likely beyond the low-energy sector usually captured by the sine–Gordon description.
In this regime, mixing between the relative and common degrees of freedom is expected to play an important role, potentially invalidating the assumption of an equilibrium sine–Gordon model~\cite{PhysRevLett.120.173601, PhysRevResearch.3.023197, PhysRevB.109.035118}.

The latent-space analysis of our VAE provides a complementary perspective (\Cref{fig:quench}b).
Before the quench, the activations are centered around positive values of $z_a$, consistent with weak interactions and large thermal fluctuations.
After the quench, rather than drifting towards negative values as one would expect from the increased coherence factor, the latent activations instead remain pinned around large positive values, even shifting slightly upward.
This behavior lies outside the range spanned by equilibrium datasets (right-most, blue points), indicating that the post-quench trajectories differ markedly from equilibrium statistics.
Inspecting the measured trajectories confirm this: large-amplitude fluctuations seeded by the initial uncoupled state persist well beyond the quench, more pronounced than in equilibrium or fast-cooled conditions (see \Cref{fig:quench_increments}).
At larger scales, the phase distribution evolves from a nearly uniform profile to one peaked around $\phi \sim 0$ with small side peaks near $\pm 2\pi$, indicating the presence of soliton-like windings similar to those observed after fast cooling.

One possible interpretation is that the system enters a \textit{prethermal} state~\cite{PhysRevLett.93.142002, doi:10.1126/science.1224953, doi:10.1126/science.1257026}. 
Prethermalization refers to the formation of a long-lived, quasi-stationary regime where a restricted set of observables rapidly relax to stationary values, while full thermalization is delayed by approximate conservation laws or weak couplings to additional degrees of freedom.
Such behavior is particularly common in nearly integrable systems, where weak integrability breaking can confine dynamics near a constrained manifold for extended times~\cite{PhysRevLett.126.090602, PhysRevX.12.041032}.
In our setting, the persistence of stable latent activations in \Cref{fig:quench}b, together with gradually evolving phase distributions, is consistent with this scenario: the relative phase field could quickly settle into a prethermal equilibrium characterized by large fluctuations and possible topological excitations, while full thermalization proceeds only slowly through coupling to the common degrees of freedom of the two superfluids~\cite{PhysRevLett.110.090404}.

At the same time, alternative explanations cannot be excluded.
The discrepancy between correlation-function analysis and latent-space signatures may also arise from experimental limitations, such as ambiguities in phase unwrapping for increments near $\Delta \phi \sim \pm \pi$, which can propagate errors through entire trajectories (see \cite{Schweigler2019} and \Cref{app:imaging}).
Although we assessed the robustness of our results by discarding problematic trajectories, such effects remain a potential source of bias.
Furthermore, given the substantial energy injected by the quench, it is plausible that the post-quench dynamics simply lie outside the sine–Gordon regime altogether, meaning that neither $M^{(4)}$ nor the latent representation can be mapped to equilibrium physics.

Regardless of the precise interpretation, our analysis highlights the complementary nature of the two approaches.
Whereas $M^{(4)}$ suggests an equilibrium-like description, the VAE exposes latent features inconsistent with equilibrium ensembles, pointing to either prethermalization or a breakdown of the sine–Gordon model.
This divergence underscores the usefulness of ML–based methods as additional probes of quantum simulators: they provide access to trajectory-level features that conventional observables may average out, thereby offering new handles on out-of-equilibrium physics.
In regards of the system studied here, and to establish unambiguous conclusions, further experiments are needed. For instance by introducing smaller quenches that inject less energy into the system, or with protocols that perform opposite quenches, from strong to weak tunnelings.
Moreover, recent techniques now permit simultaneous extraction of both relative and common phases~\cite{PhysRevResearch.7.L022031}, which would allow a direct test of whether coupling between the two sectors underlies the observed dynamics.
Extending our VAE to a multi-channel architecture could then provide a natural framework for analyzing such joint data.

\section*{Discussion}

In this work, we demonstrated that a variational autoencoder can extract a low-dimensional and physically interpretable representation of equilibrium phase trajectories from interference snapshots of tunnel-coupled 1D superfluids; a quantum simulator of the sine–Gordon field theory.
By balancing its reconstruction error with the latent space regularization, the model autonomously identified the minimal parameter set characterizing the system, namely, the dimensionless coupling $Q$, without any prior knowledge on the underlying physical model.
For this, it is crucial that the generative capabilities of the VAE’s decoder faithfully reproduce the processes’ statistics, such that the latent space correctly encodes its physical parameters and not other spurious information.
We achieved this by employing an autoregressive decoder, ensuring that the resulting phase trajectories can converge to the expected class of stochastic processes.
We showed that the decoder correctly recovers the main properties of the process, namely, its long-range correlations and non-Gaussian higher-order moments.

We then applied the equilibrium-trained model to non-equilibrium regimes.
In the case of rapid cooling, the latent representation revealed soliton defects frozen during the cooling ramp, enabling their identification at the level of individual trajectories.
In a different scenario, where the condensates undergo a sudden quench of the tunnel coupling, the VAE exposed a qualitatively distinct latent signature: despite the connected fourth moments appearing equilibrium-like, the latent activation departed strongly from the equilibrium sine–Gordon snapshots, indicating that the post-quench dynamics may lie outside the effective low-energy theory.
Together, these results show that the VAE provides a sensitive probe of both equilibrium and non-equilibrium many-body physics from limited and noisy measurements.

More broadly, this work establishes VAEs as powerful tools for the analysis of analog quantum simulation.
They address practical experimental challenges such as finite resolution, uncontrolled fluctuations, and scarce sampling, while remaining agnostic to microscopic modeling assumptions.
This capability paves the way toward data-driven discovery of new physical regimes, particularly in emerging generations of quantum simulators with unprecedented control and tunability.
While latent representations may not provide a complete description on their own, their integration with expert physical knowledge and complementary ML methods such as symbolic regression offers a scalable route toward interpretable characterization of unknown quantum dynamics.

\appendix

\section{Machine learning model}
\label{app:ml}

The architecture of our model follows closely the convolutional variational autoencoder (VAE) with an autoregressive decoder introduced in Ref.~\cite{fernandez2024learning}. For completeness, we summarize the specific design and hyperparameters used in this work.

\subsubsection{Architecture}
The model consists of three components:  
(i) an encoder that compresses each one-dimensional input trajectory of length $L=35$ into a latent representation,  
(ii) a latent space made of six probabilistic neurons $\mathbf{z} \in \mathbb{R}^6$, sampled via the reparameterization trick, and  
(iii) a decoder consisting of a convolutional upsampling stage followed by an autoregressive module.  

The encoder comprises four 1D convolutional layers with 16 channels each, followed by an adaptive pooling and a two-layer multi-layer perceptron (200 and 100 neurons). Each latent neuron is parameterized as a Gaussian distribution with mean and variance predicted by the encoder.  

The decoder mirrors the encoder structure with MLP layers (100, 200, 512), channel reshaping, interpolation, and transposed convolutions. The upsampled latent representation conditions all layers of the autoregressive module, which is implemented as a WaveNet~\cite{van2016wavenet}. This module uses a single stack of three layers of dilated convolutions with kernel size $4$ that yield a receptive field of $RF=23$. We employ zero-padding so that the autoregressive outputs have the same length as the input trajectories.

\begin{table}[h]
\centering
\begin{tabular}{ll}
\hline\hline
Layer type & Output size  \\
\hline
Input & $B \times 1 \times L$ \\
\textbf{Encoder} & \\
4 $\times$ 1D Conv. (kernel=3, stride=1) & $B \times 16 \times (L-8)$ \\
Adaptive pooling (avg + max) & $B \times 16 \times (16+16)$ \\
Flatten & $B \times 512$ \\
MLP (200, 100 neurons) & $B \times 100$ \\
Latent distribution & $B \times 12$ \\
Latent layer ($|z|=6$) & $B \times 6$ \\
\textbf{Decoder} & \\
MLP (100, 200, 512 neurons) & $B \times 512$ \\
Reshape & $B \times 16 \times 32$ \\
Interpolation & $B \times 16 \times (L-8)$ \\
3 $\times$ 1D transposed Conv.  & $B \times 16 \times L$ \\
1D transposed Conv. ($N_c=6$) & $B \times 6 \times L$ \\
WaveNet (kernel=4, layers=3) & $B \times 2 \times L$ \\
\hline\hline
\end{tabular}
\label{tab:arch}
\caption{Summary of model architecture and hyperparameters. $B = 512$ denotes the batch size. The WaveNet outputs two parameters $(\tilde\mu_j, \tilde\sigma_j)$, parameterizing a Gaussian distribution.}
\end{table}

\subsubsection{Loss function}

The model is trained by maximizing the Evidence Lower Bound (ELBO) within a Total Correlation VAE (TC-VAE) objective~\cite{Chen_Isolating_2018}. Given the increments of an input phase trajectory $\Delta\phi$, the overall loss reads
\begin{equation}
\begin{aligned}
    \mathcal{L} =& \; \mathbb{E}_q \left[ \log p_\theta(\Delta \phi | \mathbf{z}) \right] 
    - \beta \, \mathrm{KL}\!\left[q(\mathbf{z}|\boldsymbol{\phi})\|p(\mathbf{z})\right] \\
    &- \gamma \, \mathrm{TC}(\mathbf{z}) - \alpha  \, \mathrm{MI}(\mathbf{z};\boldsymbol{\phi}),
\end{aligned}
\end{equation}
with fixed parameters $\beta=3$, $\alpha = 10^{-4}$, and $\gamma = 0.1$. 

Here, each input to the model is a full wrapped phase trajectory $\boldsymbol{\phi} = (\phi_1,\dots,\phi_L)$ with $\phi_j \in [-\pi,\pi]$, while the decoder predicts the sequence of phase increments $\Delta \phi_j = \phi_j - \phi_{j-1}$.
This choice avoids issues with phase wrapping at the $\pm\pi$ boundary, since the increment variables are unbounded.
Thus, the reconstruction term is evaluated with respect to the probability distribution of increments rather than the wrapped phases themselves.

The decoder parameterizes the conditional probability of the increments in an autoregressive manner,
\begin{equation}
    p_\theta(\Delta \phi) = \prod_{j=1}^{L-1} p_\theta(\Delta \phi_j \mid \phi_{<j}, \mathbf{z}),
\end{equation}
where each conditional distribution is taken to be Gaussian,
\begin{equation}
    p_\theta(\Delta \phi_j \mid \phi_{<j}, \mathbf{z}) = 
    \mathcal{N}\!\big(\tilde\mu_j(\mathbf{z}, \phi_{<j}), \, \tilde\sigma_j^2(\mathbf{z}, \phi_{<j})\big).
\end{equation}

The reconstruction loss is therefore the negative log-likelihood (NLL) of the predicted increments,
\begin{equation}
    -\log p_\theta(\Delta \phi | \mathbf{z}) 
    = - \sum_{i=1}^N \sum_{j=1}^{L-1} 
    \log \mathcal{N}\!\left(\Delta \phi_j^{(i)} \;\middle|\; 
    \tilde\mu_j^{(i)}, \tilde\sigma_j^{2(i)}\right),
\end{equation}
which, for a single data point, expands to
\begin{equation}
    -\log \mathcal{N}(\Delta \phi_j \mid \tilde\mu_j, \tilde\sigma_j^2) 
    = \log \!\left(\tilde\sigma_j \sqrt{2\pi}\right) 
    + \frac{(\Delta \phi_j - \tilde\mu_j)^2}{2 \tilde\sigma_j^2}.
\end{equation}

The minimum of this loss is achieved when the predicted mean $\tilde\mu_j$ matches the observed increment $\Delta \phi_j$, while a smaller predicted variance $\tilde\sigma_j^2$ reduces the loss further. In this way, the network learns to output both the most likely increment and an uncertainty estimate that reflects the stochastic nature of the experimental trajectories. The TC and MI penalty terms act to encourage disentangled, minimally correlated latent variables, while the KL term aligns the approximate posterior with the unit Gaussian prior.

\subsubsection{Training}
The model is trained on experimental equilibrium trajectories across several tunnel couplings. Each training consists of 128 epochs with a maximum learning rate of $5 \times 10^{-4}$, optimized using Adam~\cite{kingma2014adam} and the one-cycle learning rate~\cite{smith2018disciplined,smith2018super} scheduler provided by \texttt{fastai}~\cite{howard2020fastai}. Batch size is $B = 512$. All model parameters are initialized using Kaiming initialization~\cite{he2015delving} (fan-out mode), with latent variances initialized to zero.

\section{Transfer matrix formalism and It\^{o} process for the sine--Gordon model}
\label{app:ito}

\begin{figure}
    \centering
    \includegraphics{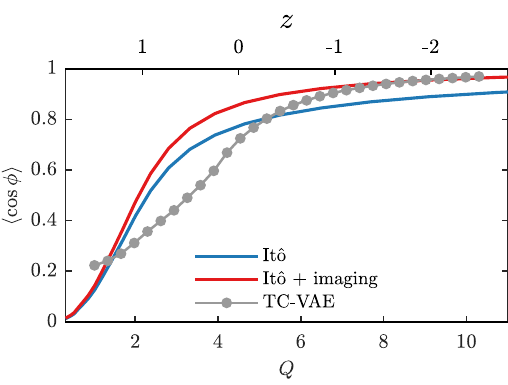}
    
    \caption{\label{fig:coherence_Ito}
    Coherence factor $\langle \cos \phi \rangle$ as function of effective interaction strength $Q$ of the sine--Gordon model. Phase trajectories are sampled using the It\^{o} process~\Cref{eq:ito}. The experimental imaging process is modeled by convolving $\phi$ with a Gaussian, then coarse-graining.
    On the top x-axis, the coherence factor as function of the active latent variable $z$ is plotted for phase trajectories generated with the trained model.
    }
\end{figure}

\begin{figure*}
    \centering
    \includegraphics{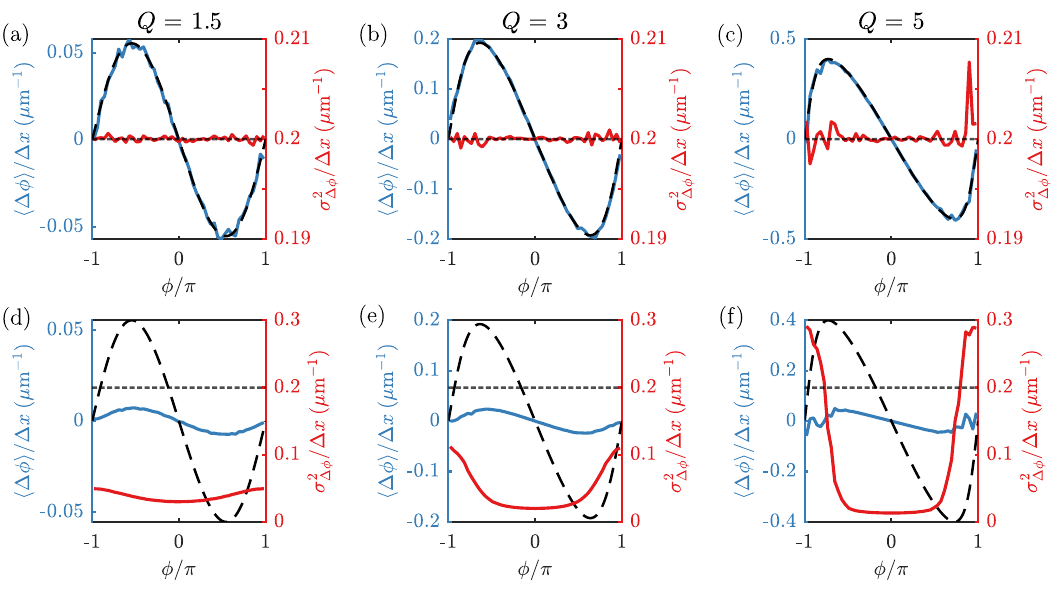}
    
    \caption{\label{fig:Ito_phase_increments}
    Phase increments from the It\^{o} process of \Cref{eq:ito}. The left vertical axis shows the mean phase increment $\langle \Delta \phi \rangle$ as a function of the instantaneous phase $\phi$ (solid blue), compared to the theoretical drift term $A$~\Cref{eq:Ito_drift} (long-dashed black). The right vertical axis shows the standard deviation of the phase increments (solid red), together with the theoretical diffusion coefficient $2D = 4/\lambda_T$ (short-dashed black). Panels (a–c) (top row) present results obtained directly from the It\^{o} process for three values of the effective interaction strength $Q$. Panels (d–f) (bottom row) show the corresponding results after accounting for imaging effects, implemented by convolving the phase trajectories with a Gaussian point-spread function to model finite resolution, followed by coarse-graining. 
    }
\end{figure*}

The stochastic It\^{o} process used to sample equilibrium phase trajectories of the sine--Gordon model [Eq.~\eqref{eq:ito}] follows from the transfer matrix formalism (see, e.g., Refs.~\cite{schweigler2017experimental, PhysRevA.98.023613}).
Here we briefly summarize the steps specific to the sine--Gordon case, referring to the general derivation to these references.  

For two tunnel--coupled one-dimensional superfluids, the classical Hamiltonian reduces under the standard approximations (decoupling of relative and common degrees of freedom) to the sine--Gordon Hamiltonian given in the main text.
Within the transfer matrix formalism, thermal correlation functions are generated by the auxiliary operator
\begin{equation}
    \hat{K}_{\mathrm{SG}}= \frac{1}{\lambda_T} \left(-2 \partial_\phi^2-2 \frac{Q^2}{8} \left( \cos \phi -1 \right)\right) \, ,
\end{equation}
where $\lambda_T = 2 \hbar^2 n / (m k_B T)$ is the thermal coherence length and $Q$ the effective interaction strength.  
The equilibrium distribution of relative phases is determined by the ground state $\Psi_0(\phi)$ of $\hat{K}_{\mathrm{SG}}$, which obeys a Mathieu-type equation.

By general arguments (see Ref.~\cite{risken1989fpe}), the corresponding Fokker--Planck equation is equivalent to an It\^{o} process of the form
\begin{equation}
    \mathrm{d}\phi = A[\phi]\, \mathrm{d}x + \sqrt{2 D\, \mathrm{d}x}\; \mathcal{N}(0,1) \, ,
\end{equation}
with diffusion constant $D = 2/\lambda_T$.  
The deterministic drift is given by
\begin{equation}
    A[\phi] = - 2D \, \frac{\partial}{\partial \phi} \ln | \Psi_0(\phi) | \, ,
    \label{eq:Ito_drift}
\end{equation}
which encodes the effect of the tunnel coupling through the cosine potential.  

Thus, equilibrium phase trajectories of the sine--Gordon model can be generated efficiently by integrating the It\^{o} process using the drift determined from $\Psi_0(\phi)$ and the universal diffusion constant set by $\lambda_T$.
In practice, we use numerical solutions of the Mathieu equation to evaluate $\Psi_0(\phi)$ and hence the drift $A[\phi]$~\cite{PhysRevA.87.013629}.

\section{Imaging effects and simplified modeling}
\label{app:imaging}

\begin{figure*}
    \centering
    \includegraphics{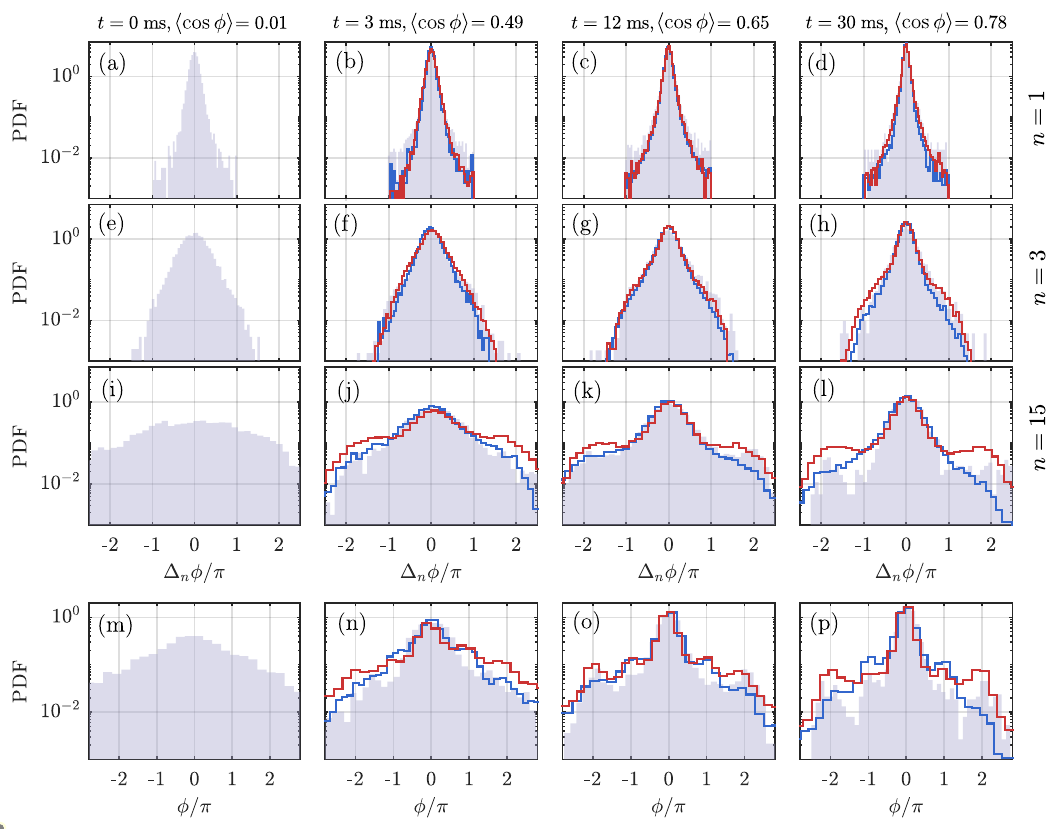}
    
    \caption{\label{fig:quench_increments}
    Distributions of phase increments and phase values following quench of the tunnel coupling $J$ (purple shaded) compared to slow- (blue lines) and fast-cooled (red lines) data at similar coherence factor $\langle \cos \phi \rangle$.
    The phase increments are computed at various distances, such that $\Delta_{n}\phi_i = \phi_i - \phi_{i-n}$.
    (a-d) Phase increments between neighboring points $n =1$.
    (e-h) Phase increments at 3 pixels separation $n=3$.
    (i-l) Phase increments at 15 pixels separation $n= 15$.
    (m-p) Distributions of phase values $\phi$.
    }
\end{figure*}

The relative phase field $\phi(x)$ is experimentally accessed through matter-wave interference imaged on a CCD camera with finite pixel size $\Delta x \approx 2~\mu$m. This process inevitably distorts the underlying phase fluctuations of the condensates. Two main effects are relevant:
(i) the imaging system has a finite point spread function (PSF), effectively a Gaussian blurring with width $\sigma \sim 3~\mu$m, which smooths the phase trajectory over neighboring points;
(ii) coarse-graining occurs when the CCD integrates over the size of the individual pixels, further reducing resolution at length scales smaller than $\Delta x$. Together, these effects suppress short-wavelength fluctuations of the phase field, modify the measured correlation functions, and can mix drift and diffusion contributions of the It\^{o} process for the sine--Gordon phase field~\Cref{eq:ito}, leading to an apparent phase-dependent diffusion term (see \Cref{fig:latent_analysis}d).

A comprehensive treatment of the imaging process has been developed in Ref.~\cite{PhysRevA.104.043305} and in the detailed discussion of Ref.~\cite{Schweigler2019}, where the full absorption imaging dynamics, integration along the imaging axis, and technical noise sources are modeled.
Such approaches are necessary for precise quantitative comparison of microscopic observables, in particular when extracting parameters like the effective diffusion constant from first principles.

For the purposes of the present work, however, a simplified but well-controlled approximation is sufficient. We account for imaging resolution by convolving the generated phase trajectories with a Gaussian of width $\sigma$ corresponding to the effective PSF, followed by coarse-graining to the experimental pixel size $\Delta x$.
This procedure reproduces the main observable consequences of imaging: the change of the coherence factor is modest (\Cref{fig:coherence_Ito}), but higher-order quantities, such as the variance of increments, are more strongly affected.
In particular, the convolution mixes neighboring phases such that the measured It\^{o} process acquires an effective $\phi$-dependent diffusion (see \Cref{fig:Ito_phase_increments}), consistent with the trends seen in experiment.

Because our VAE is trained directly on experimental trajectories, which already include these imaging effects, the simplified model is only used when making equilibrium checks against theory (e.g., It\^{o} sampling).
In these cases, the Gaussian-convolution approach captures the essential impact of imaging without requiring the full machinery of absorption-imaging simulations.

Additional effects/defects may arise when the relative phase is extracted from the absorption imaging (see Ref.~\cite{Schweigler2019} for full details). 
To extract the phase, the interference pattern is fitted with a cosine function with a Gaussian envelope, from which the phase modulo $2 \pi$ is extracted. 
In order to ensure a continuous phase trajectory, it is assumed that the phase difference between neighboring pixels does not exceed $\pi$.
However, in the presence of large fluctuations, found, for instance, following the quench of the tunnel coupling, errors in the extracted phase trajectory may occur.
In \Cref{fig:quench_increments}, distributions of the phase increments before and following the quench of the tunnel coupling are shown; indeed, no increments with amplitude greater than $\pi$ is found.
Comparing with phase increment distributions of the slow and fast cooled data, we do observe a slightly larger number of increments near the values $\phi = \pm 1$, which could indicate an increased number of errors in the extracted phase trajectories.
Nevertheless, compared to the total number of increments, the number of potential errors is rather low. 
Indeed, filtering out profiles containing increments $| \Delta \phi | \geq 0.9 \pi$ does not significantly change our results.

\newpage

\bibliography{references}

\section*{Acknowledgments}
This research was funded in part by the European Research Council: ERC Advanced Grant "Emergence in Quantum Physics" (EmQ) under Grant Agreement No. 101097858, and ERC Advanced Grant "Artificial agency and learning in quantum environments
" (QuantAI) under Grant Agreement No. 101055129. This work was also supported by the Austrian Science Fund (FWF) [SFB BeyondC F7102, 10.55776/F71]. For open access purposes, the author has applied a CC BY public copyright license to any author accepted manuscript version arising from this submission.
G.F-F. acknowledges the European Research Council AdG NOQIA; MCIN/AEI (PGC2018-0910.13039/501100011033, CEX2019-000910-S/10.13039/501100011033, Plan National FIDEUA PID2019-106901GB-I00, Plan National STAMEENA PID2022-139099NB, I00, project funded by MCIN/AEI/10.13039/501100011033 and by the ``European Union NextGenerationEU/PRTR’’ (PRTR-C17.I1), FPI); QUANTERA DYNAMITE PCI2022-132919 under Grant Agreement No. 101017733; Ministry for Digital Transformation and of Civil Service of the Spanish Government through the QUANTUM ENIA project call - Quantum Spain project, and by the European Union through the Recovery, Transformation and Resilience Plan - NextGenerationEU within the framework of the Digital Spain 2026 Agenda; Fundació Cellex; Fundació Mir-Puig; Generalitat de Catalunya (European Social Fund FEDER and CERCA program); Barcelona Supercomputing Center MareNostrum (FI-2023-3-0024); (HORIZON-CL4-2022-QUANTUM-02-SGA PASQuanS2.1, 101113690, EU Horizon 2020 FET-OPEN OPTOlogic, Grant No. 899794, QU-ATTO, 101168628), EU Horizon Europe Program (This project has received funding from the European Union’s Horizon Europe research and innovation program under grant agreement No. 101080086 NeQST); ICFO Internal ``QuantumGaudi’’ project;
 The views and opinions expressed in this article are however those of the author(s) only and do not necessarily reflect those of the European Union or the European Research Council - neither the European Union nor the granting authority can be held responsible for them.
\end{document}